\documentclass[3p,times]{elsarticle}
\usepackage{graphics}
\usepackage{eurosym}
\usepackage{amsfonts}
\usepackage{amssymb}
\usepackage{amsmath}
\usepackage{booktabs}

\usepackage{graphicx}
\usepackage{graphicx,epsfig}
\usepackage[font={footnotesize,it}]{caption}
\usepackage[colorlinks=false,
linkcolor=red,
urlcolor=red,
citecolor=blue]{hyperref}

\usepackage{float}
\usepackage{color}
\usepackage{mathptmx}
\usepackage{anyfontsize}
\usepackage{t1enc}
\usepackage{amssymb}

\usepackage[figuresright]{rotating}

\begin{document}

\begin{frontmatter}




\title{A generalized buchdahl model for compact stars in $f(R,T)$ gravity}


\author{J. Kumar}
\ead{jitendark@gmail.com}
\author{H. D. Singh}
\ead{drdipalisingh25@gmail.com}
\author{A. K. Prasad}
\ead{amitkarun5@gmail.com}

\address{Department of Applied Mathematics, Central University of Jharkhand, Ranchi-835205, India}

\begin{abstract}
In this paper, we study the stellar structure in terms of alternative theory of gravity specially by $f(R,T)$ gravity theory. Here, we consider the function $f(R,T)=R+2\varsigma T$ where $R$ is the Ricci scalar, $T$ is the stress-energy momentum and $\varsigma$ is the coupling constant. Using it we developed a stellar model that briefly explains the isotropic matter distribution within the compact object filled with perfect fluid. The stability of the model is shown by several physical and stability conditions. With the accecptibility of our theory, we were able to collect data  for compact stars like PSR-B0943+10, CEN X-3, SMC X-4, Her X-1 and 4U1538-52 with great accuracy.

\end{abstract}


\end{frontmatter}


\section{Introduction}
Recently, it has been discovered that our universe was expanding at an accelerated rate. Over past few years, more-n-more studies were done to understand these late-time acceleration of the universe and evidences were found that the cause of this is due to the presence of some dark-energy matter \cite{1,2,3,4,5,6,7} which has huge negative impact and pressure. Surveys on cosmic radiation and dark matters \cite{8,9,10,11} shown that $4\%$, $20\%$ and $76\%$ of ordinary baryonic matter, dark matter and dark energy are distributed respectively over the universe. As the gravitational connection to this is least understood and Einstein's general theory on relativity was unable to provide decent reasons behind such causes.\par
So, different approaches were made for investigating such cases. And one approach is the consideration of modified theory of relativity which is a classical generalization of general relativity that were able to explain late-time acceleration and early-time inflation along with dark-energy. Suggestion were made that by replacing/ substituting curvature invariant function in Einstein-Hilbert action we can achieve cosmic acceleration. Many studies were done in past years on modified gravity like $f(R)$, $f(G)$ or $f(R,G)$ in different contexts \cite{12,13,14,15,16,17,18,19,20,21}. Similarly, another modified theory is $f(R,T)$ theory of gravity \cite{22}. In this theory, gravitational Lagrangian couples with trace $T$ of the energy-momentum tensor of the matter is added to general Einstein-Hilbert action. Now, with its introduction, many researchers tried their best in providing information and genuine explaination of various cosmological and astrophysical events \cite{23,24,25,26,27,28,29,30,31,32}. In paper \cite{33}, $f(R,T)$ has been used in the studies of factors that affect the stability of isotropic spherical symmetry. And by applying perturbation scheme \cite{34,35,36} instability of spherically symmetric under different condition was studied. In a recent aricle, stellar model in the framework of $f(R,T)$ gravity has been studied \cite{37} that shows this model has more impressive behaviour when compared to the general relativity theory. In another paper \cite{38}, the effect of modified $f(R,T)$ has been studied and compared with the results of standard general relativity. Futher, investigation was done by applying Krori and Barua metric \cite{39}. In article \cite{40}, they have distinguish between general relativity and its extended version for relativistic stars which has firm gravitational field system.\par 
However, in this paper we did an investigation of stellar model in the reference of $f(R,T)$ behaviour using Buchdahl metric for different compact stars. First of all, for a given star we determined the energy density and isotropic pressure in the context of $f(R,T)$, then the velocity of sound and Energy condition (weak, null and strong). Later, we further  investigated the Redshift and also the Adiabatic Index of the star. This paper is arranged in the following sequences : In sect 1, we have given brief introduction regarding General Relativity and Modified Theory of Relativity. In sect 2, we have presented the basic formulation of $f(R,T)$ theory of gravity. In sect 3, we have shown stellar structure in the framework of $f(R,T)$ Theory. In sect 4, we did a detail investigation of compact stars using Buchdahl metric for cases $K<0$ and $K>1$. Further in sect 5 and 6, we did observation regarding the physical quantities and stabilities of compact stars. Finally, in sect 7, we discussed and concluded our results. 

\label{}
\section{ General mathematical formulation of $f(R,T) $ gravity theory} 
In this section, we introduced the formation of $f(R,T)$ \cite{41} gravity theory. While obtaining the Einstein's field equations, the ricci scalar in Einstein-Hilbert action  is integrated over $d^4 x$ as 
\begin{eqnarray}
	S_{EH}=\frac{1}{16}\int{R\sqrt{-g}d^4 x}\label{1}
\end{eqnarray} 
Now one can get $f(R,T)$ field equation from the above equation if instead of Ricci scalar $R$, we select $f(R,T)$. Thus, the full action for the modified theory of gravity is given by 
\begin{eqnarray}
	S=\frac{1}{16}\int{f(R,T)\sqrt{-g}d^4 x}+\int{L_M \sqrt{-g}d^4 x}\label{2}
\end{eqnarray} 
where in the first term $ f(R,T) $ is arbitrary function of Ricci scalar $R$ with $g$ as the determinant of the metric tensor $g_{\mu\nu}$ and $T$ being the trace of the stress-energy tensor,$T_{\mu\nu}$. And in the second term  $L_M$ is the matter Lagrangian density which is related to stress-momentum tensor given as 
\begin{eqnarray}
	T_{\mu\nu}=-\frac{2}{\sqrt{-g}} \frac{\delta(\sqrt{-g} L_M)}{\delta g^{\mu \nu}} \label{3}
\end{eqnarray}
with trace $T=g^{\mu\nu}T_{\mu\nu}$. Assuming that the Lagrangian density $L_M$ \cite{42} only depends on the metric tensor components $g_{\mu\nu}$, we obtain eqn.(\ref{3}) as  
\begin{eqnarray}
	T_{\mu\nu}=g_{\mu\nu}L_M-\frac{2\delta( L_M)}{\delta g^{\mu \nu}} \label{4}
\end{eqnarray}
By varying eqn.(\ref{2}) with respect to $g_{\mu\nu}$, we obtained the field equation 
\begin{align}
	(R_{\mu\nu}-\bigtriangledown_{\mu}\bigtriangledown_{\nu})f_{R}(R,T)+g_{\mu\nu}\diamondsuit  f_{R}(R,T)-\frac{1}{2}f(R,T)g_{\mu\nu} \nonumber\\
	=& 
	8\pi T_{\mu\nu}-f_{T}(R,T)(T_{\mu\nu}+\varPsi_{\mu\nu}) \label{5}
\end{align}	
where $f_R(R,T)=\frac{\delta f(R,T)}{\delta R}$ and $f_T(R,T)=\frac{\delta f(R,T)}{\delta T}$. The covariant derivative is denoted by $\bigtriangledown_{\mu}$ and the diamond operature $\diamondsuit$ is defined as 	 
\begin{eqnarray*}
	\diamondsuit=\frac{1}{\sqrt{-g}} \frac{\delta}{\delta x^{\mu}}(\sqrt{-g} g^{\mu\nu} \frac{\delta}{\delta x^{\nu}})
\end{eqnarray*}

\begin{eqnarray*}
	\varPsi_{\mu\nu}=g^{\alpha\beta}\frac{\delta T_{\alpha\beta}}{\delta g^{\mu\nu}}
\end{eqnarray*}
Now we perform the covariant derivative of eqn.(\ref{5}) by which we obtain the covariant derivative of the energy-momentum tensor as 
\begin{eqnarray}
	\bigtriangledown^{\mu} T_{\mu\nu}=\frac{f_T(R,T)}{8\pi-f_T(R,T)} [ (T_{\mu\nu}+\varPsi_{\mu\nu})\bigtriangledown^{\mu}ln f_T(R,T)+\varPsi_{\mu\nu}\bigtriangledown^{\mu}-\frac{1}{2}g^{\mu\nu}\bigtriangledown^{\mu} T ]  \label{6}
\end{eqnarray}
Clearly from eqn.(\ref{6}) we can see that the stress-energy momentum tensor $T_{\mu\nu}$ in $f(R,T)$ theory does not follow the conservation law as in Einstein general relativity (GR) due to the presence of some nonminimal matter geometry coupling in its formulation. By using eqn.(\ref{4}), the tensor $\varPsi_{\mu\nu}$ is given by
\begin{eqnarray}
	\varPsi_{\mu\nu}=-2T_{\mu\nu}+g_{\mu\nu}L_M -2g^{\alpha\beta}\frac{\delta^2 L_M}{\delta g^{\mu\nu}\delta g^{\alpha\beta}} \label{7}
\end{eqnarray} 
To find the field equations, we consider the interior of star to be filled with a perfect fluid source along with the energy-momentum tensor of the form
\begin{eqnarray}
	T_{\mu\nu}=(\rho+p)u_{\mu}u_{\nu}-pg_{\mu\nu}\label{8}
\end{eqnarray}   
provided the four velocity $u_{\nu}$ satifies $u_{\mu}u^{\mu}=-1$ and $u_{\nu}\bigtriangledown^{\mu} u_{\nu}=0$, $\rho$ and $ p $ is the matter density and isotropic pressure respectively . Taking the matter Lagrangian as $L_M=-p$, we obtain eqn.(\ref{7}) as 
\begin{eqnarray}
	\varPsi_{\mu\nu}=-2T_{\mu\nu}-pg_{\mu\nu} \label{9}
\end{eqnarray}
In this paper, we assumed the function $f(R,T)$as $f(R,T)=R+2f(T)$ where $f(T)$ is the function of trace $(T)$ of the stress-energy tensor of matter. Here, we choose $f(T)=\varsigma T$ in order to determine the modified theory of gravity, where $\varsigma$ is a coupling constant. \\
Using the expression in eqn.(\ref{5}), we obtain 
\begin{eqnarray}
	G_{\mu\nu}=8\pi T_{\mu\nu}+\varsigma Tg_{\mu\nu}+2\varsigma(T_{\mu\nu}+pg_{\mu\nu}) \label{10}
\end{eqnarray}
When $f(R,T)\equiv R$, then eqn.(\ref{5}) reduces to Einstein field equations. By putting $f(R,T)=R+2\varsigma T$ and eqn.(\ref{9}) in eqn.(\ref{6}) we get
\begin{eqnarray}
	\bigtriangledown^{\mu}T_{\mu\nu}=-\frac{\varsigma}{2(4\pi+\varsigma)}[g_{\mu\nu}\bigtriangledown^{\mu}T+2\bigtriangledown^{\mu}(pg_{\mu\nu})]\label{11}
\end{eqnarray}
Thus, one can easily obtain the conservation equation in Einstein's gravity by putting $\varsigma=0$.
For studying cosmological and astrophysical problems $f(R,T)$ has been widely considered due to its advantages in explaining such space problems.

\section{ Einstein's field equation in $f(R,T)$ gravity theory}
Defining the interior space-time of static stellar configuration for spherically symmetric metric, by the line element as
\begin{eqnarray}
	ds^{2}=-e^{\lambda(r)}dr^{2} -r^{2}(d\theta^{2}+\sin^{2}\theta d\phi^{2})+e^{\nu(r)}dt^{2}\label{12}
\end{eqnarray} 
where $(t,r,\theta,\phi)$ are the time and space cordinates respectively and $\lambda(r)$ and $\nu(r)$ are the functions of radial cordinate, $r$ only.\\
Now, we have to find the field equations of eqn.(\ref{12}) as \cite{43}  
\begin{eqnarray}
	\rho_{E} =\frac{e^{-\lambda}}{8\pi} \bigg(-\frac{1}{r^2}+\frac{\lambda^{\prime}}{r}+\frac{e^{\lambda}}{r^2} \bigg) \label{13}\\
	p_{E}=\frac{e^{-\lambda}}{8\pi} \bigg(\frac{1}{r^2}+\frac{\nu^{\prime}}{r}-\frac{e^{\lambda}}{r^2} \bigg)\label{14} \\
	p_{E}=\frac{e^{-\lambda}}{32\pi}\bigg(2\nu''+\nu'^2-\lambda'\nu'+2\frac{\nu'-\lambda'}{r}\bigg)\label{15}
\end{eqnarray}
where $\rho_{E}$ and $p_{E}$ are the Einstein's density and pressure respectively and prime refers to differentiation with respect to $r$. Along with

\begin{eqnarray}
	\rho_{E}=\rho_{F}+\frac{\varsigma}{8\pi}(3\rho_{F}-p_{F}) \nonumber \\
	p_{E}=p_{F}+\frac{\varsigma}{8\pi}(\rho_{F}-3p_{F}) \nonumber
\end{eqnarray}
where $p_{F}$ and $\rho_{F}$ denotes the $f(R,T)$ pressure and density respectively. Here, for our convenience, we have assumed $G=c=1$. 
On solving eqns.(\ref{13})-(\ref{15}), we obtain
\begin{eqnarray}
	\frac{\nu'}{2}(p_{F}+\rho_{F})+\frac{dp}{dr}=\frac{\varsigma}{8\pi+2\varsigma}(p'-\rho')\label{16}
\end{eqnarray}
which provides the hydrostatic equillibrium condition for general relativity when we take $\varsigma=0$.\\
Now rewriting modified pressure and density i.e $p_{F}$ and $\rho_{F}$ by using eqn.(\ref{13}) and eqn.(\ref{14}), we get
\begin{eqnarray}
	p_{F}=\frac{\pi}{(8\pi^2+6\pi \varsigma +\varsigma^2)}[(8\pi+3\varsigma) p_{E}+\varsigma \rho_{E}] \label{17}\\
	\rho_{F}=\frac{\pi}{(8\pi^2+6\pi \varsigma +\varsigma^2)}[(8\pi+3\varsigma)\rho_{E}+\varsigma p_{E}] \label{18}
\end{eqnarray}
Now, it has been assumed that all the cosmological phenomena are engulfed in the vacuum space-time, so we need to check our solution for interior space-time with the exterior Schwarzschild line element as
\begin{eqnarray}
	ds^2=-\bigg(1-\frac{2M}{r}\bigg)dt^2+\frac{dr^2}{1-\frac{2M}{r}}+r^2 \bigg(d\theta^2+sin^2\theta d\phi^2\bigg)\label{19}
\end{eqnarray}
where $M$ is the total mass with radius $R$. At boundary when $r=R$
\begin{eqnarray}
	e^{-\lambda(R)}=e^{\nu(R)}=1-\frac{2M}{R}\label{20}
\end{eqnarray}
Now, we need to solve the field equation i.e eqn.(\ref{17}) and eqn.(\ref{18}) for which we approached with Buchdahl ansatz and study its impact on the physical systems.

\section{ Buchdahl ansatz in $f(R,T)$ gravity theory}
In order to find the solution of the modified field equation, we use Buchdahl anastz \cite{44} which easily investigate all the physical parameters of a stellar model. The most broadly studied Buchdahl metric anastz is given by
\begin{eqnarray} 
	e^{\lambda}=\dfrac{K(1+Cr^{2})}{K+Cr^{2}} \label{21}
\end{eqnarray} 
where $K$ and $C$ are constant parameters that depends on the geometry of the given star.
\ \ Now we consider few transformations for solving field equations ,
\begin{eqnarray} 
	\chi=\sqrt{\frac{K+Cr^2}{K-1}}~~~~  \textrm{and}~~~~~ e^{\nu}=W^2 \label{22} 
\end{eqnarray}
Substituting eqn.(\ref{21}) and eqn.(\ref{22}) in eqn.(\ref{13}) and eqn.(\ref{14}) and doing some simple algebraic calculations we obtained the following equations which will be later used in solving exact solutions, 
\begin{align}
	-\kappa p_{E} &=\frac{(K+Cr^2)}{K(1+Cr^2)}\bigg[-\frac{2W'}{rW}+\frac{C(K-1)}{K+Cr^2}\bigg] \label{24} \\
	-\kappa p_{E} &=\frac{(K+Cr^2)}{K(1+Cr^2)}\bigg[-\frac{W''}{W}+\frac{W'}{rW}-\frac{C(K-1)}{(K+Cr^2)(1+Cr^2)}\bigg(1+\frac{rW'}{W}\bigg)\bigg] \label{25} \\
	\kappa c^2 \rho_{E} &= \frac{C(K-1)(3+Cr^2)}{K(1+Cr^2)^2} \label{26}
\end{align}
Now subtracting eqn.(\ref{25}) from eqn.(\ref{24}), we get
\begin{eqnarray}
	\dfrac{K+Cr^{2}}{K(1+Cr^{2})}\Bigg[\dfrac{W''}{W}-\dfrac{W'}{rW}+\dfrac{C(K-1)r(Cr-W'/W)}{(K+Cr^{2})(1+Cr^{2})}\Bigg]=0 \label{27}
\end{eqnarray}
Using  eqn.(\ref{22}) in the above equation, we obtained the differential equation of the form,
\begin{eqnarray}
	(1-\chi^2)\frac{d^2 W}{d\chi^2}+\chi \frac{dW}{d\chi}+(1-K)W=0 \label{28}
\end{eqnarray}
To find the solution of the above equation we use transformation for two cases :\\
Case 1 for $K<0$ given as
\begin{eqnarray}
	W=(1-\chi^{2})^{1/4}U\label{29}
\end{eqnarray}
Thus eqn.(\ref{28}) reduces to
\begin{eqnarray}
	\dfrac{d^{2}U}{dX^{2}}+\psi U=0\label{30}
\end{eqnarray}
where
\begin{eqnarray} \psi=\frac{1-K}{1-\chi^2}-\frac{2+3\chi^2}{4(1-\chi^2)^2} \nonumber
\end{eqnarray}
In order to solve the eqn.(\ref{30}) easily,  we choose $ \psi $ as
\begin{eqnarray}
	\psi=\frac{A_1}{\chi^2}\label{31}
\end{eqnarray}
where $A_1$ be any arbitrary constant.\\
\ \ \ Thus eqn.(\ref{30}) becomes 
\begin{eqnarray}
	\dfrac{d^{2}U}{dX^{2}}+\frac{A_1}{\chi^2} U=0\label{32}
\end{eqnarray}
Solving the differential equation we obtain the solution as
\begin{eqnarray}
	U=c_1 \frac{\chi^{1-A_1}}{1-A_1}+c_2 \nonumber
\end{eqnarray}
Therefore, we get
\begin{eqnarray}
	W=(1-\chi^2)^{1/4} \bigg[c_1 \frac{\chi^{1-A_1}}{1-A_1}+c_2 \bigg]\label{33}
\end{eqnarray}
where $c_1$ and $c_2$ are arbitrary constants.\\
\ \ \ Using eqn.(\ref{33}) in eqns.(\ref{24}) and (\ref{26}) we obtained equations,
\begin{align}
	p_E &=\frac{c}{8\pi K(1+Cr^2)}\bigg[ -\frac{\chi^2}{1-\chi^2}-(K-1)+\frac{2\chi^{1-A_1}}{\bigg[ \frac{\chi^{1-A_1}}{1-A_1}+\frac{c_2}{c_1}\bigg]}\bigg] \label{34} \\
	\rho_E &=\frac{C(K-1)(3+Cr^2)}{8\pi K(1+Cr^2)^2} \label{35}
\end{align}
Substituting the above equations in eqns.(\ref{17})-(\ref{18}), we obtained the required modified pressure and density, respectively
\begin{align}
	p_{F} &=\frac{\pi}{(8\pi^2+6\pi \varsigma +\varsigma^2)} \bigg[ \frac{(8\pi+3\varsigma)}{8\pi K(1+Cr^2)}\bigg[ -\frac{\chi^2}{1-\chi^2}-(K-1)+ \nonumber\\  &\ \ \ \ \frac{2\chi^{1-A_1}}{\bigg[ \frac{\chi^{1-A_1}}{1-A_1}+\frac{c_2}{c_1}\bigg]}\bigg]+\varsigma \frac{(K-1)(3+Cr^2)}{8\pi K(1+Cr^2)^2}\bigg] \label{36}\\
	\rho_{F} &=\frac{\pi}{(8\pi^2+6\pi \varsigma +\varsigma^2)} \bigg[(8\pi+3\varsigma) \frac{(K-1)(3+Cr^2)}{8\pi K(1+Cr^2)^2}+\frac{\varsigma }{8\pi K(1+Cr^2)} \nonumber\\  &\ \ \ \ \bigg[  -\frac{\chi^2}{1-\chi^2}- (K-1)+ \frac{2\chi^{1-A_1}}{\bigg[ \frac{\chi^{1-A_1}}{1-A_1}+\frac{c_2}{c_1}\bigg]}\bigg] \label{37}
\end{align}

\begin{figure}[h]
	\includegraphics[width=4cm]{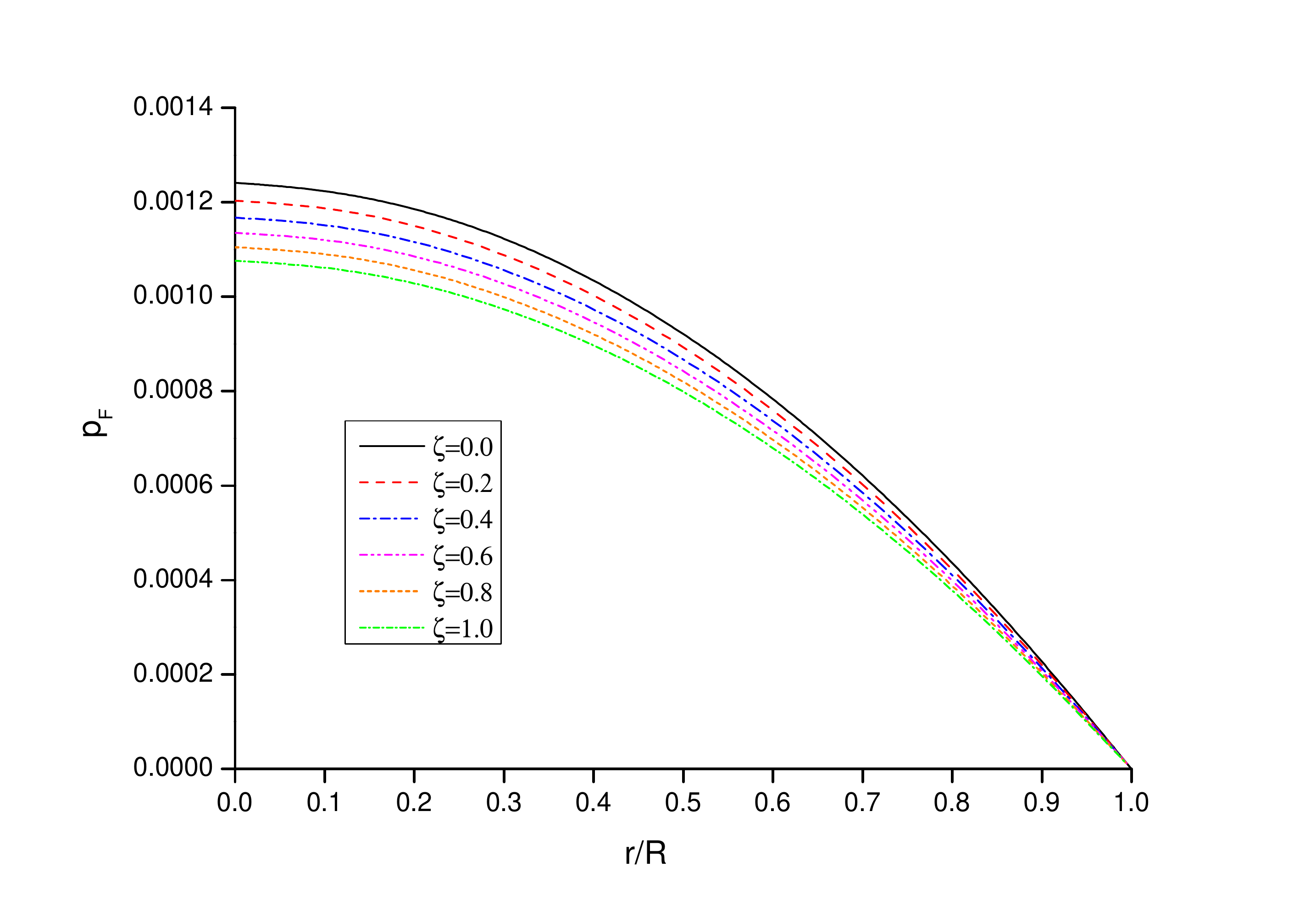}\includegraphics[width=4cm]{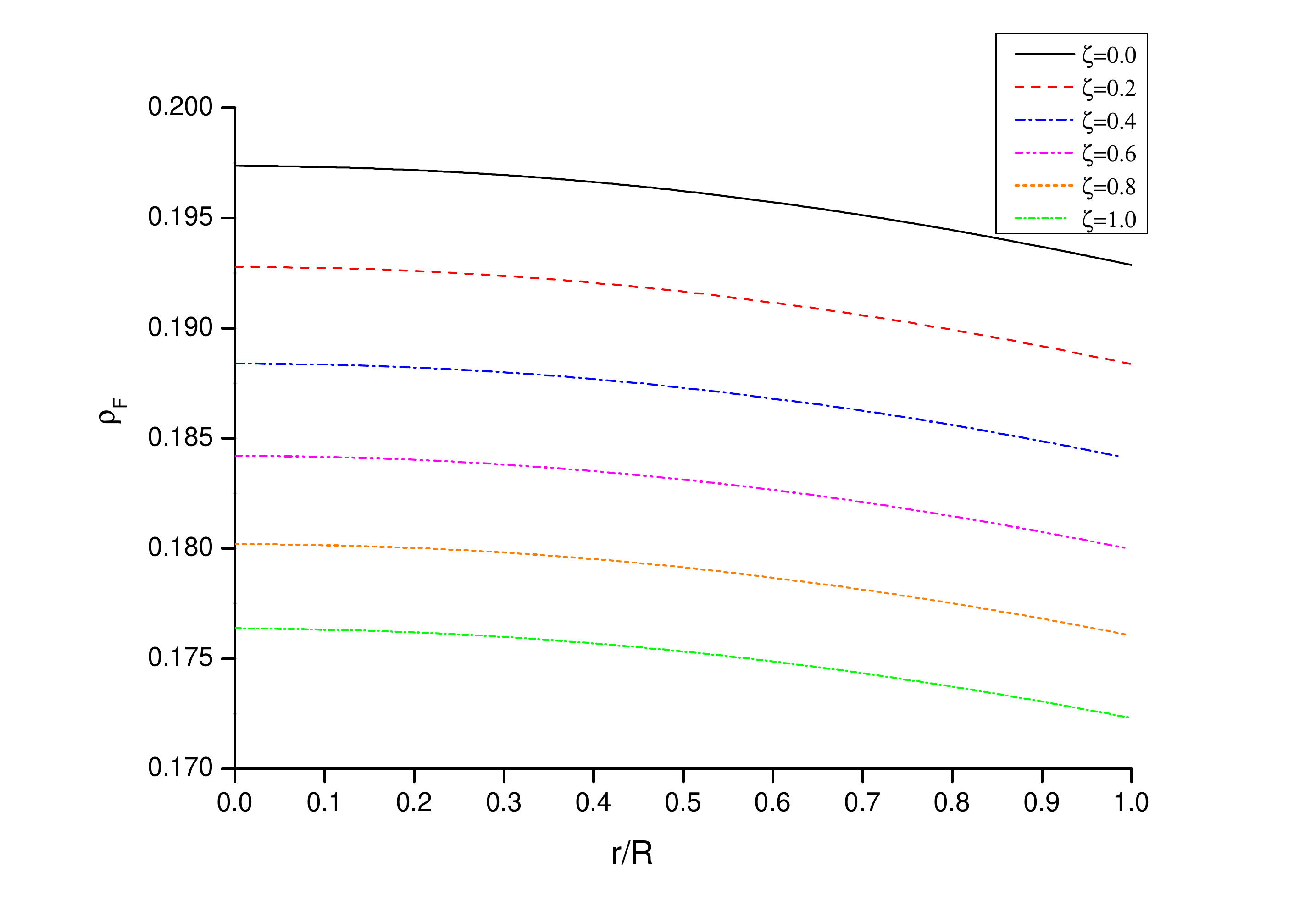}	
	\includegraphics[width=4cm]{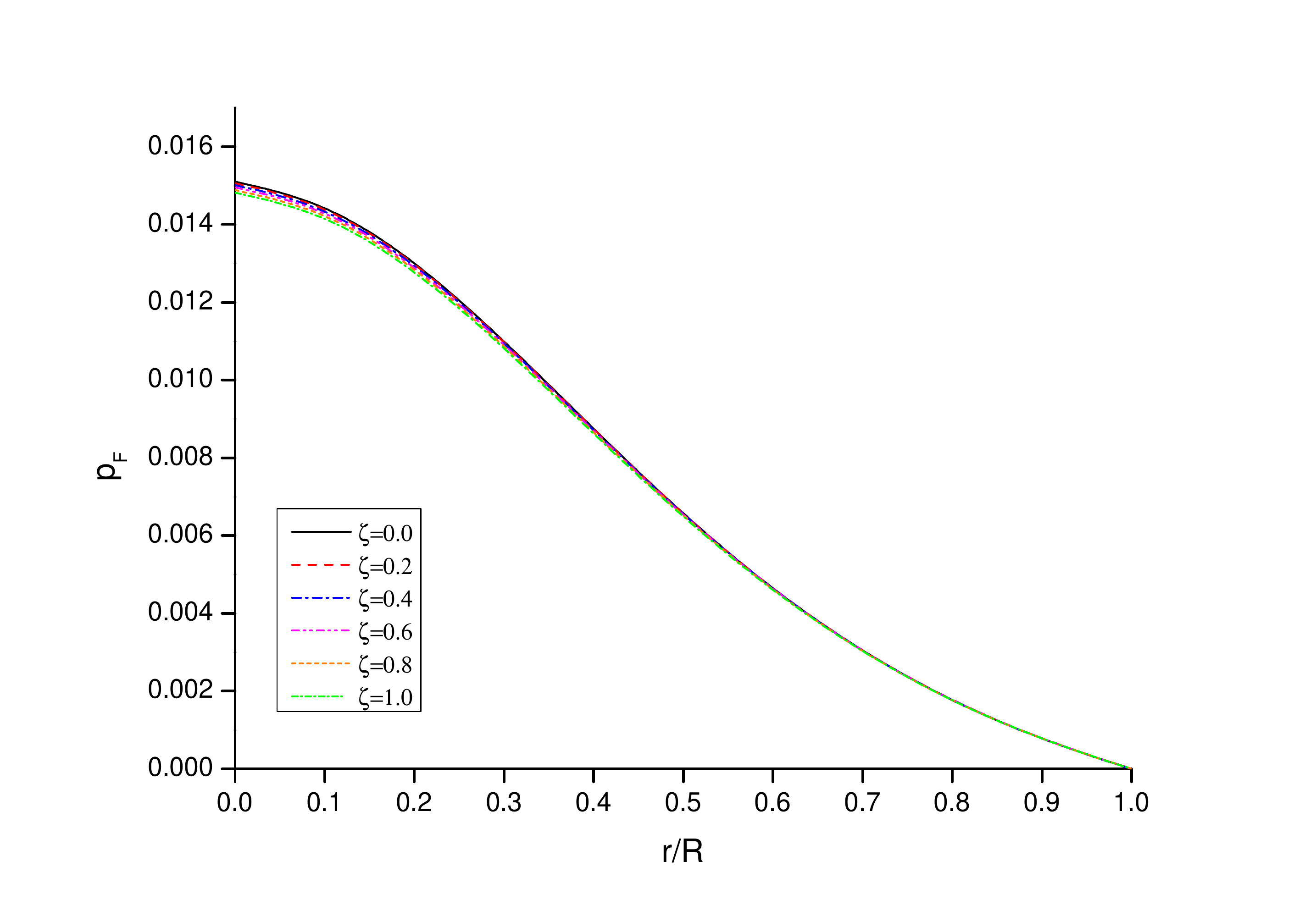}\includegraphics[width=4cm]{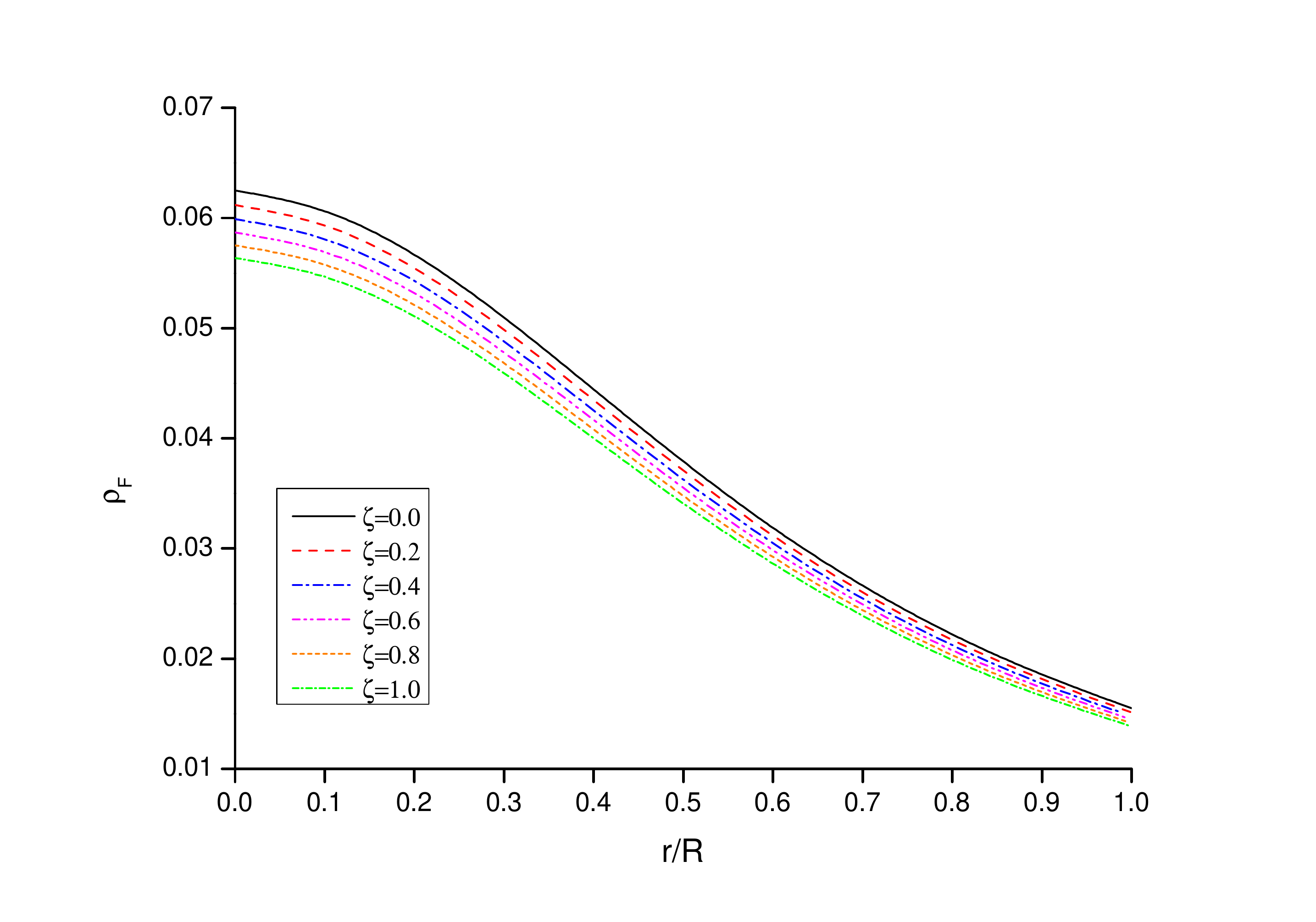}
	\caption{Above two graphs represent the behaviour of pressure and Denity against fractional radius $r/R$ of  compact star PSR B0943+10 for $K=-1.53$, $C=0.002071 Km^{-2}$, $A_1=0$, $M=0.02M_{\odot}$, $R=2.6 Km$ where as the below two graphs represent compact star Her X-1 for $K=2.1$, $C=0.022018 Km^{-2}$, $A_1=0$, $M=0.85M_{\odot}$, $R=8.1 Km$  \label{f1}}
\end{figure}

Case 2 for $K>1$ , we consider the transformation,
\begin{eqnarray}
	W=(\chi^2 -1)^{1/4} \bigg[c_1 \frac{\chi^{1-A_1}}{1-A_1}+c_2 \bigg]\label{38}
\end{eqnarray}
where $ c_1 $ and $ c_2 $ are arbitrary constants. Here, we find the pressure and density is same as for $K<0$. When $0<K<1$ , it is observed that due to the presence of the term $K-1$  the density becomes negative. Where as for pressure the transformation $\chi$ has no real solution in $0<K<1$. Hence, for given value of $K$, physical validation for general relativity and modified theory is not possible. Now, the interior of a stellar model has to follow some standard restrictions as follows which can be seen in {fig \ref{f1}}.
\begin{enumerate}
	\item Pressure and density should be positive definite at the centre.
	\item The pressure should be maximum at the centre and decreasing monotonically within $0<r<R$.
	\item The density should be maximum at the centre and decreasing monotonically within $0<r<R$.
	\item The ratio of pressure and density should be less than unity within $0<r<R$.
\end{enumerate}

\section{Determination of constant}
Using eqn.(\ref{20}) with boundary condition i.e when $r=R$ and $p_{F}=0$, we obtained the value of the arbitary constant as 
\begin{eqnarray}
	F=\frac{M1 (8\pi+3\varsigma) (2\chi^{1-A_1})(8\pi K(1+CR^2)^2)}{[M1 (8\pi+3\varsigma) (8\pi K(1+CR^2)^2)](K-1+\chi^2)-M2} - M3 \label{39}
\end{eqnarray}
where $F=\frac{c_2}{c_1}$, $M1=\frac{1}{8\pi K(1+CR^2)}$, $M2= \varsigma (K-1)(3+CR^2)$ and $M3=\frac{\chi^{1-A_1}}{1-A_1}$. 

\section{Physical properties to be satisfied by the stellar structure of $f(R,T)$ gravity theory}
To show the effectiveness of the our stellar model, we studied and analyzed some physical properties and depicted it graphically. Here, we study the Velocity of Sound, Energy Conditions, Redshift, Adiabetic Index etc. in the following subsections. 

\subsection{\textit{Velocity of Sound}}
\begin{figure}[h]
	\includegraphics[width=6cm]{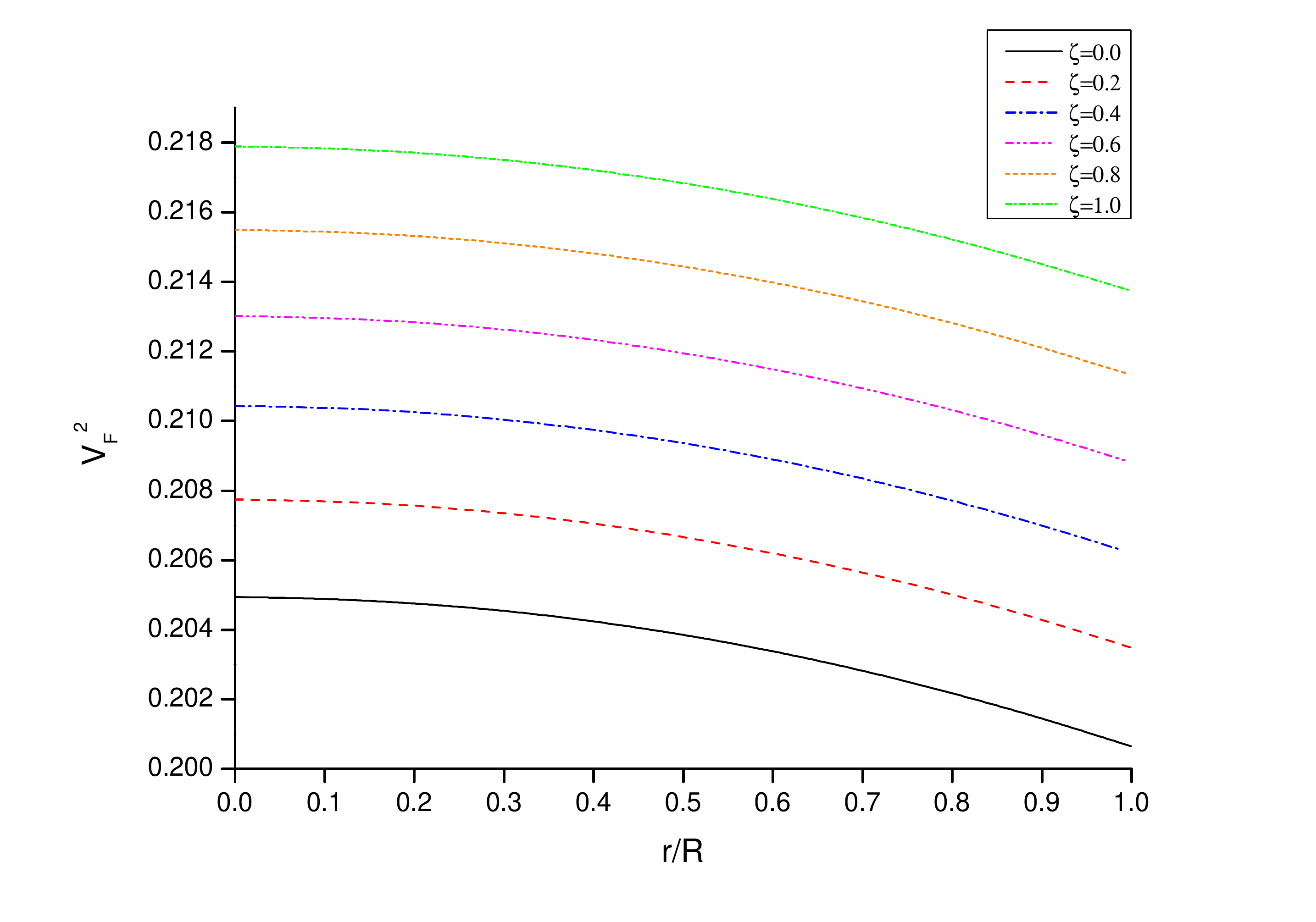}\includegraphics[width=6cm]{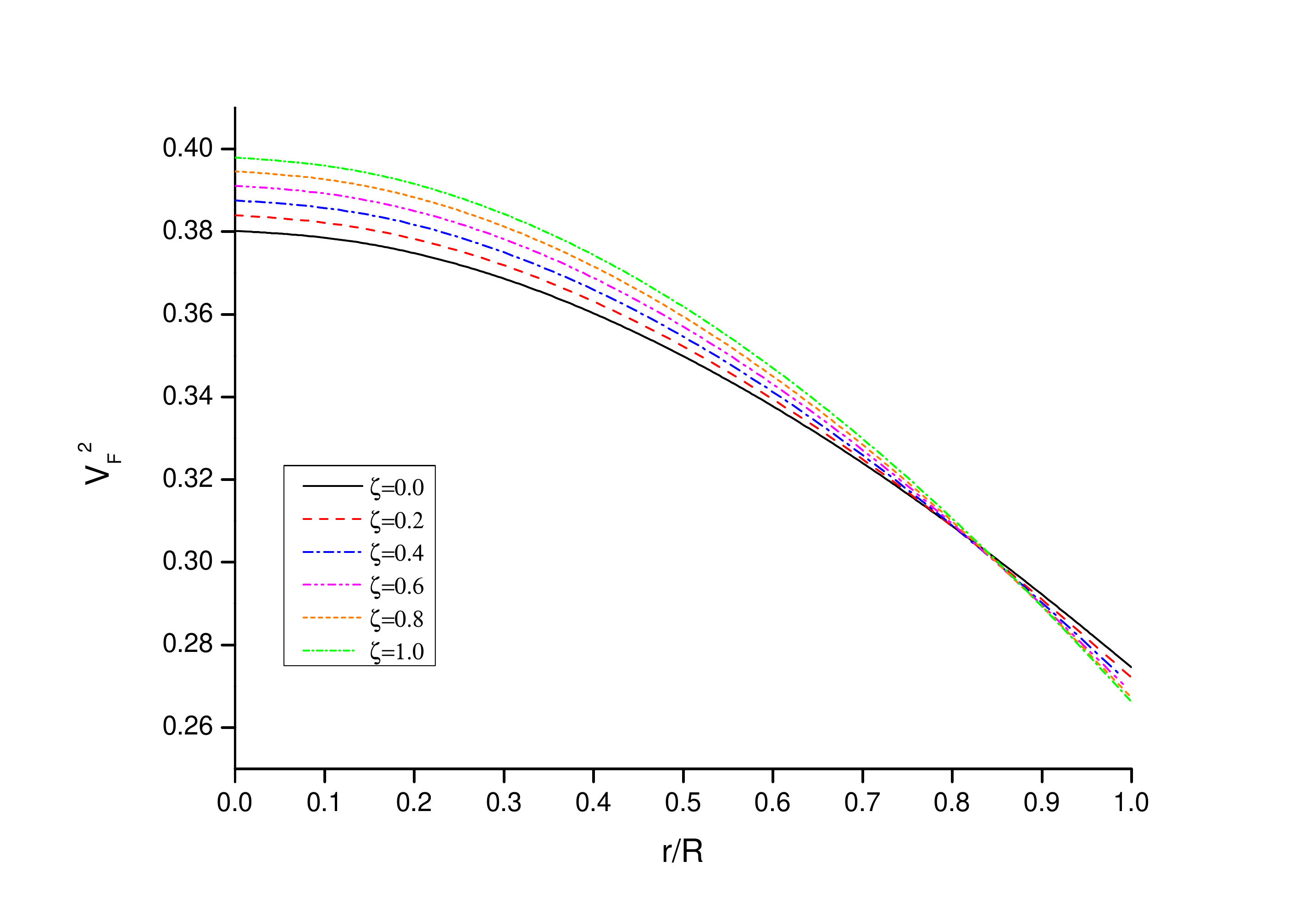}
	\caption{Left and Right graphs represents the behaviour of Velocity of sound against fractional radius $r/R$ of compact stars PSR B0943+10 and Her X-1 respectively.   \label{f2}}	
\end{figure}

In order to establish the stability of the stellar system, it must satisfy the causality condition which is the velocity of sound should be less than the velocity of light i.e, $0< V^2 < 1$.\\
\ \ \ \ The velocity of sound is obtained as 
\begin{eqnarray}
	V^2_{F}=\frac{dp_{F}}{d \rho_{F}} \label{40}
\end{eqnarray}
In {fig \ref{f2}}, shows that for the stellar structure the velocity of sound is less than the speed of light inside compact stars and with increase in $r$, it decreases.

\subsection{\textit{Redshift}}
The expression for gravitational Redshift $Z_{S}$ for the stellar model is given by 
\begin{eqnarray}
	Z_S=e^{-\nu(r)/2}-1 \label{41}
\end{eqnarray}

\begin{figure}[h]
	\includegraphics[width=6cm]{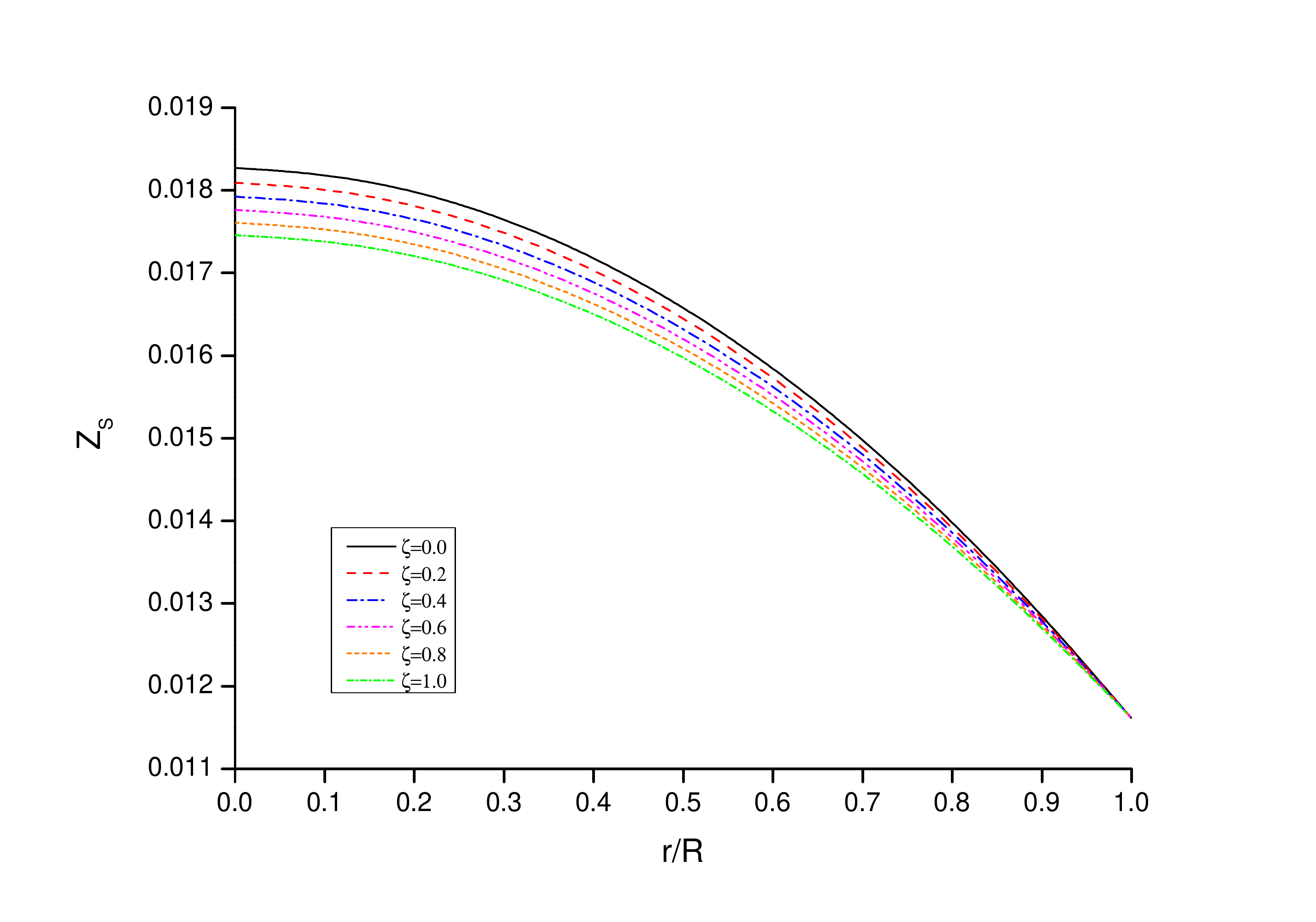}\includegraphics[width=6cm]{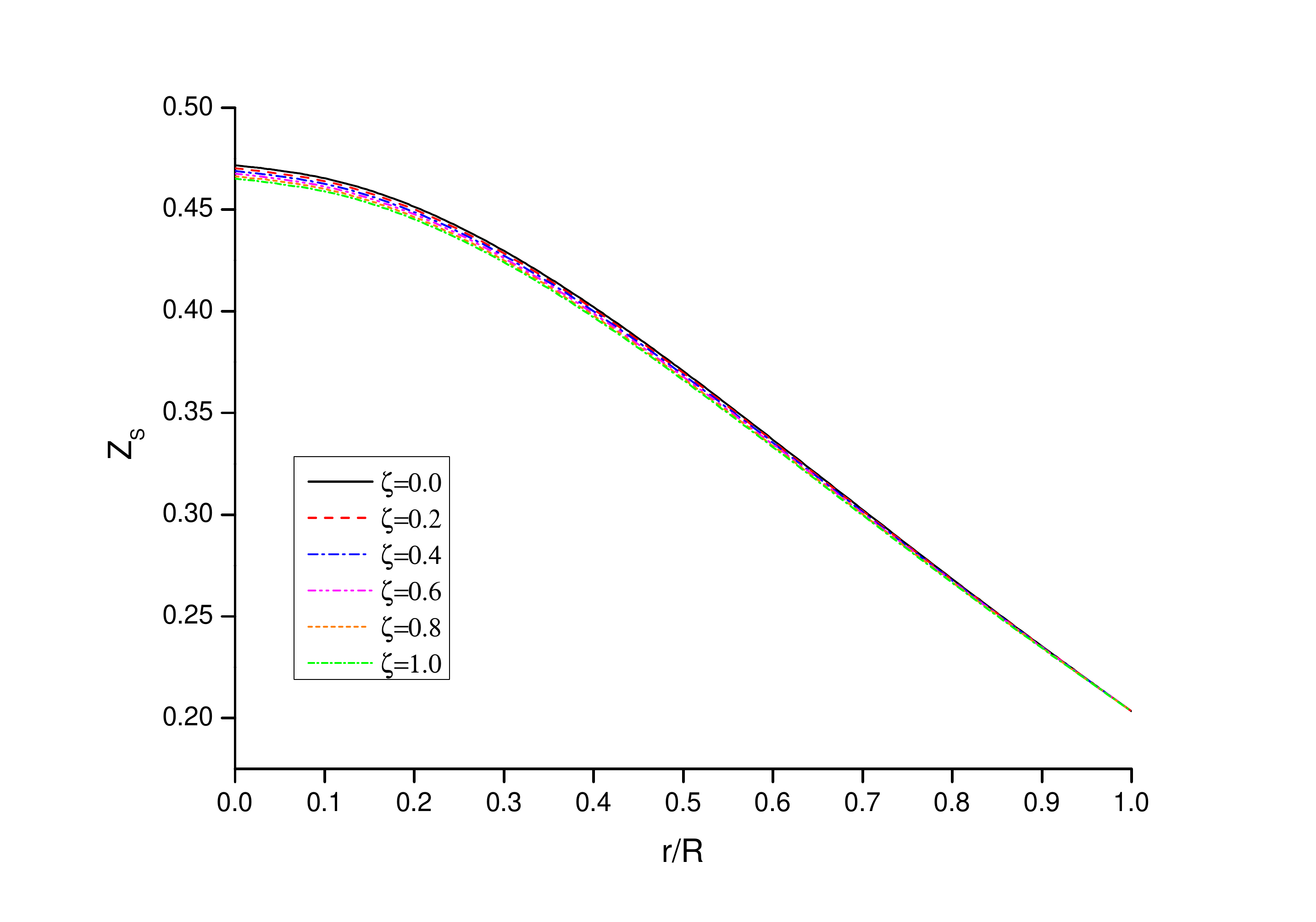}
	\caption{Behaviour of Redshift against fractional radius $r/R$ of compact stars PSR B0943+10 and Her X-1 respectively. \label{f4}}	
\end{figure}

which should be positive inside the astrophysical objects. In {fig \ref{f4}}, we have shown the gravitational redshift inside the compact stars which indicate the physical validity of the matter distribution of our stellar model. 
\subsection{\textit{Energy Condition}}
\begin{figure}[h]
	\includegraphics[width=5.5cm]{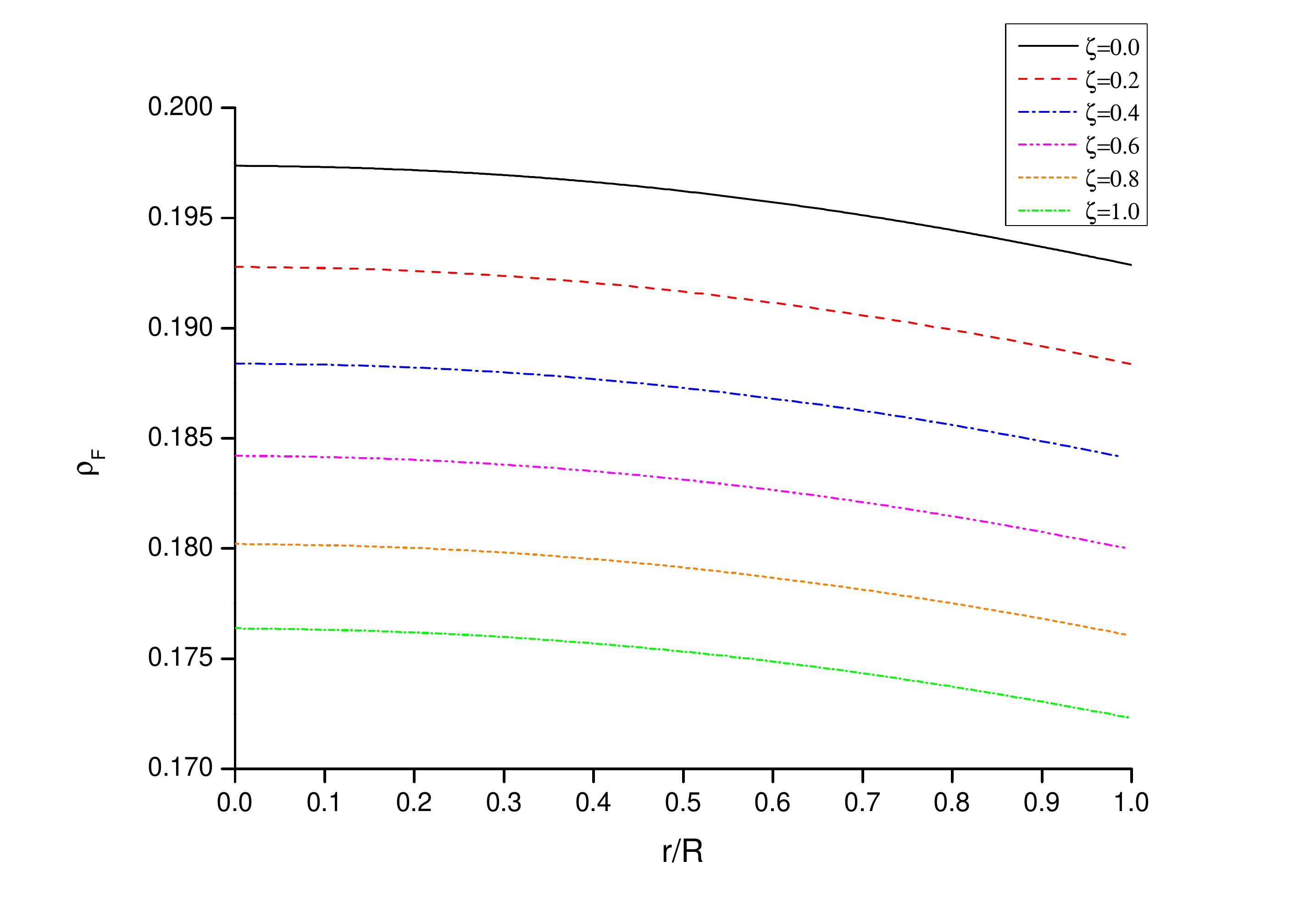}\includegraphics[width=5.5cm]{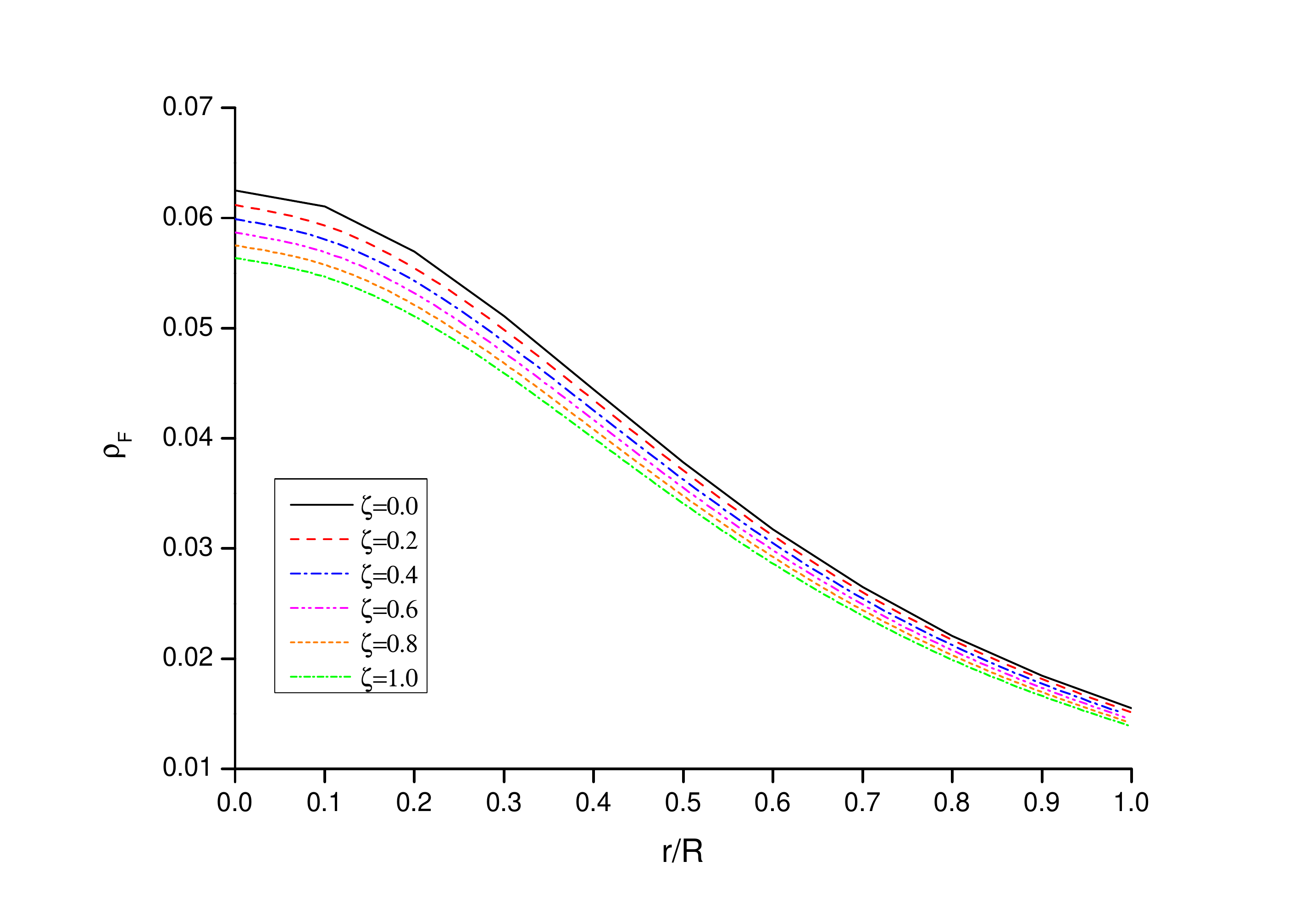}\includegraphics[width=5.5cm]{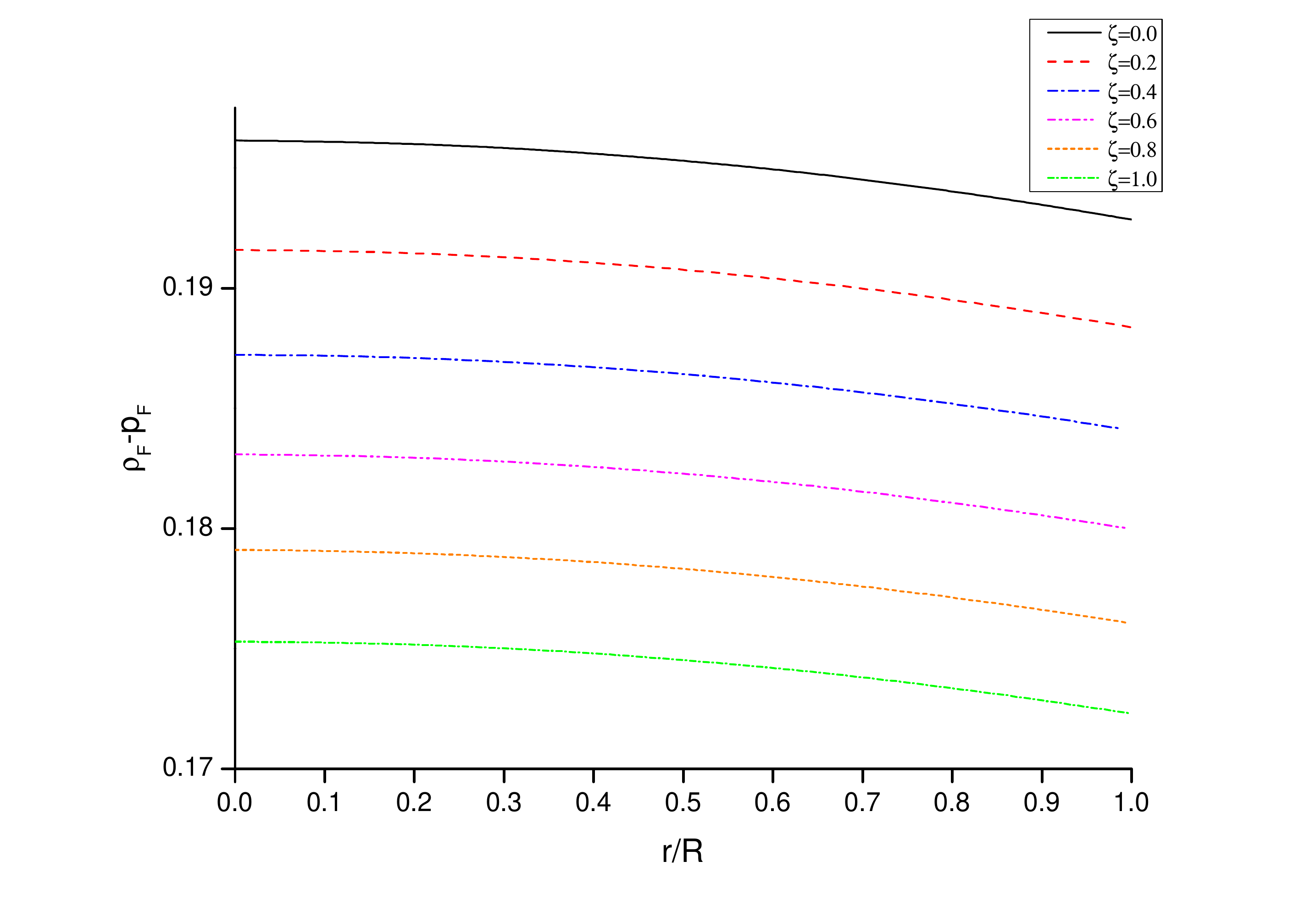}
	\includegraphics[width=5.5cm]{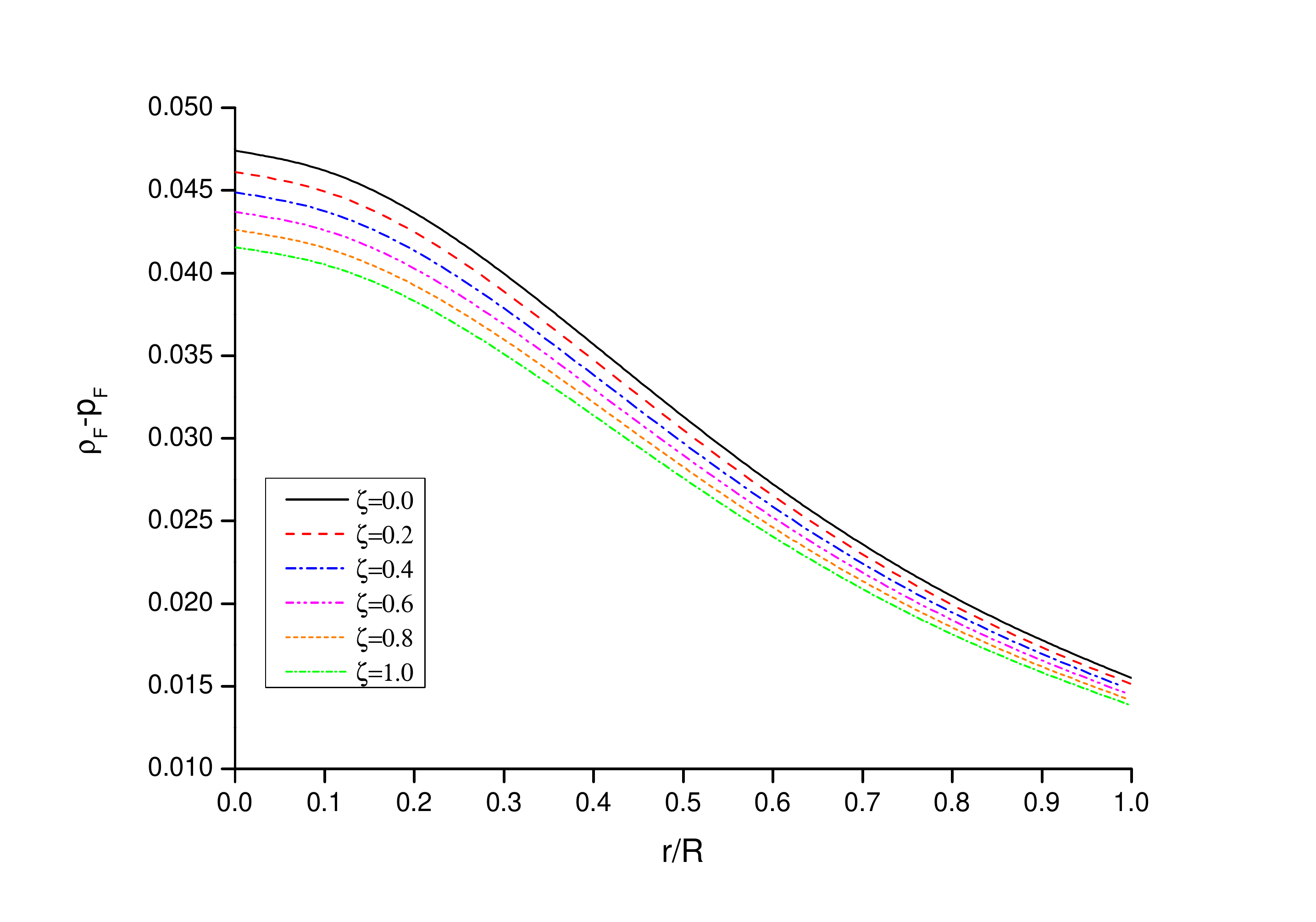} \includegraphics[width=5.5cm]{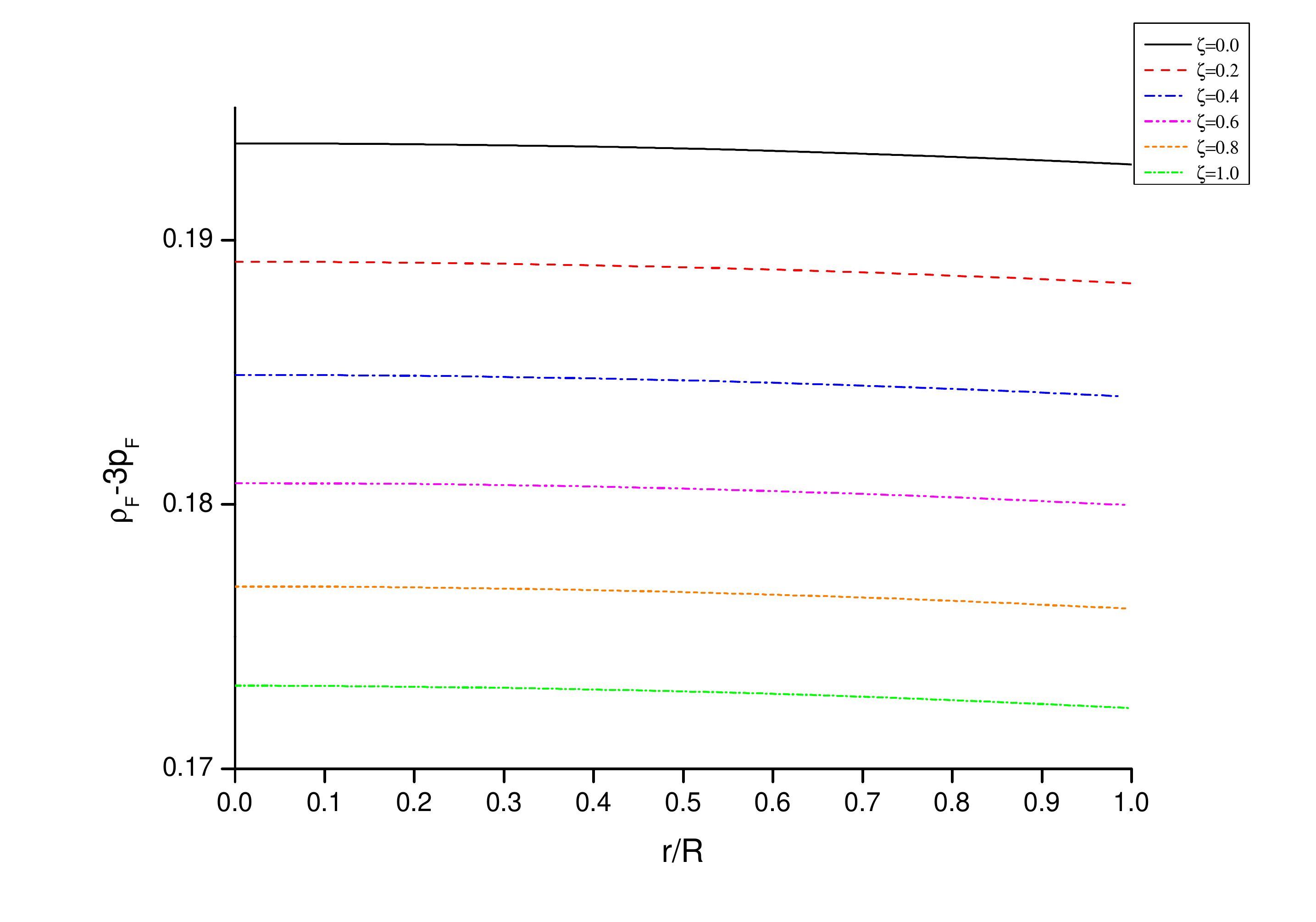}\includegraphics[width=5.5cm]{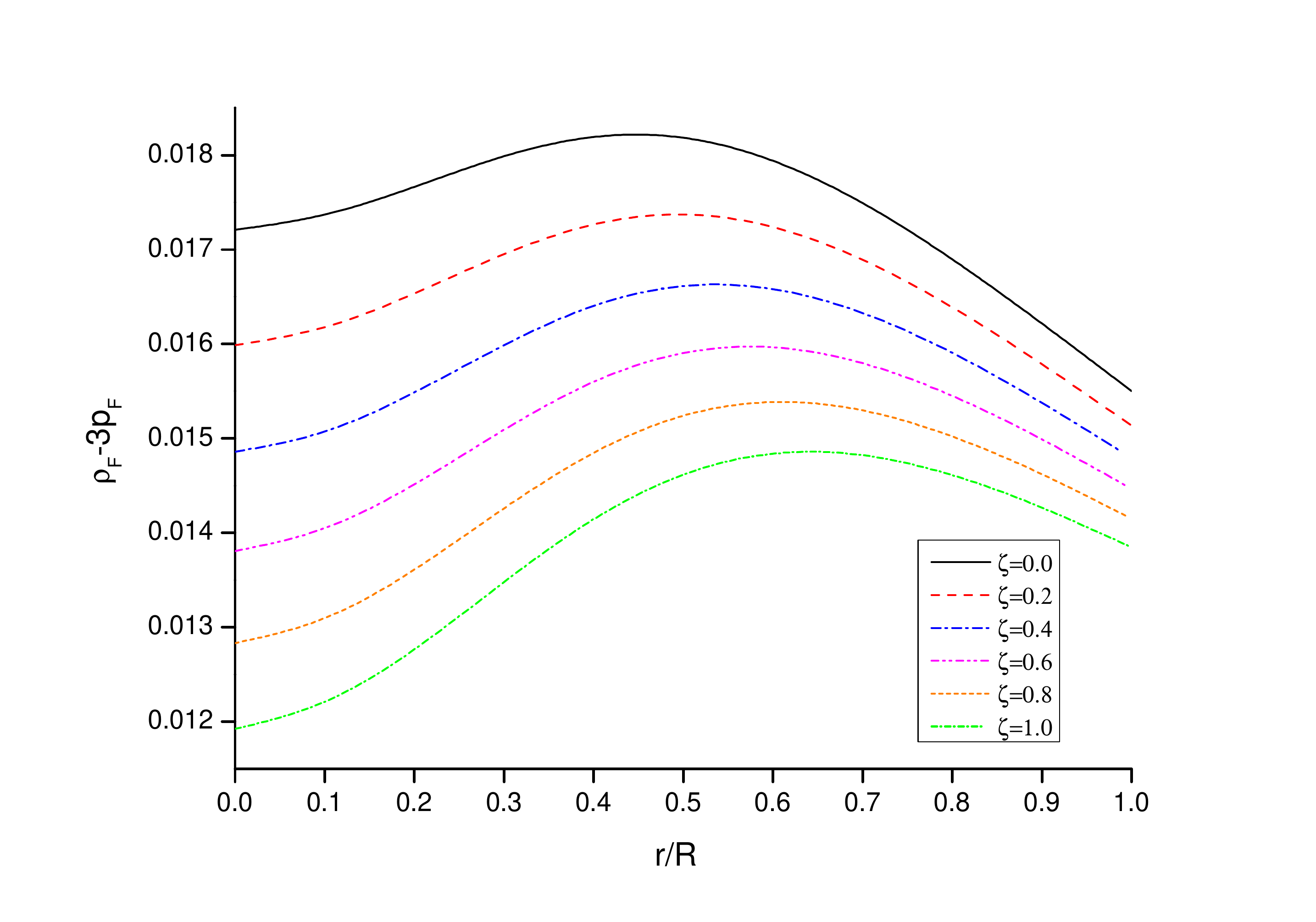}
	\caption{Top graphs represent the Null energy condition, whereas the middle graphs represent the Weak energy condition and  the end graphs represent the Strong energy condition  of compact stars PSR B0943+10 and Her X-1 respectively.  \label{f3}}	
\end{figure}

In order to show the consistency of the stellar system, it must satisfy the energy condition as given  in the following inequalities,
\begin{align}
	& Null Energy Condition: & NEC= &\rho_{F} \geq 0 \nonumber\\
	& Weak Energy Condition: & WEC= &\rho_{F} - p_{F} \geq 0 \nonumber\\
	& Strong Energy Condition:& SEC= &\rho_{F} - 3p_{F} \geq 0 \nonumber
\end{align}
In {fig \ref{f3}}, we can see the behaviour of these energy conditions and observed that throughout the stellar model all the energy conditions are satisfied and evenly distributed over the interior region of the compact stars. 

\subsection{\textit{Adiabatic Index}}
In order to investigate the stability of the model, we have to study the adiabatic index $\Gamma_I$ of the system, which is defined by
\begin{eqnarray}
	\Gamma_I=\bigg(\frac{c^2 \rho_F + p_F}{p_F}\bigg) \bigg(\frac{dp_F}{c^2 d \rho_F}\bigg)\label{42}
\end{eqnarray} 

\begin{figure}[h]
	\includegraphics[width=6cm]{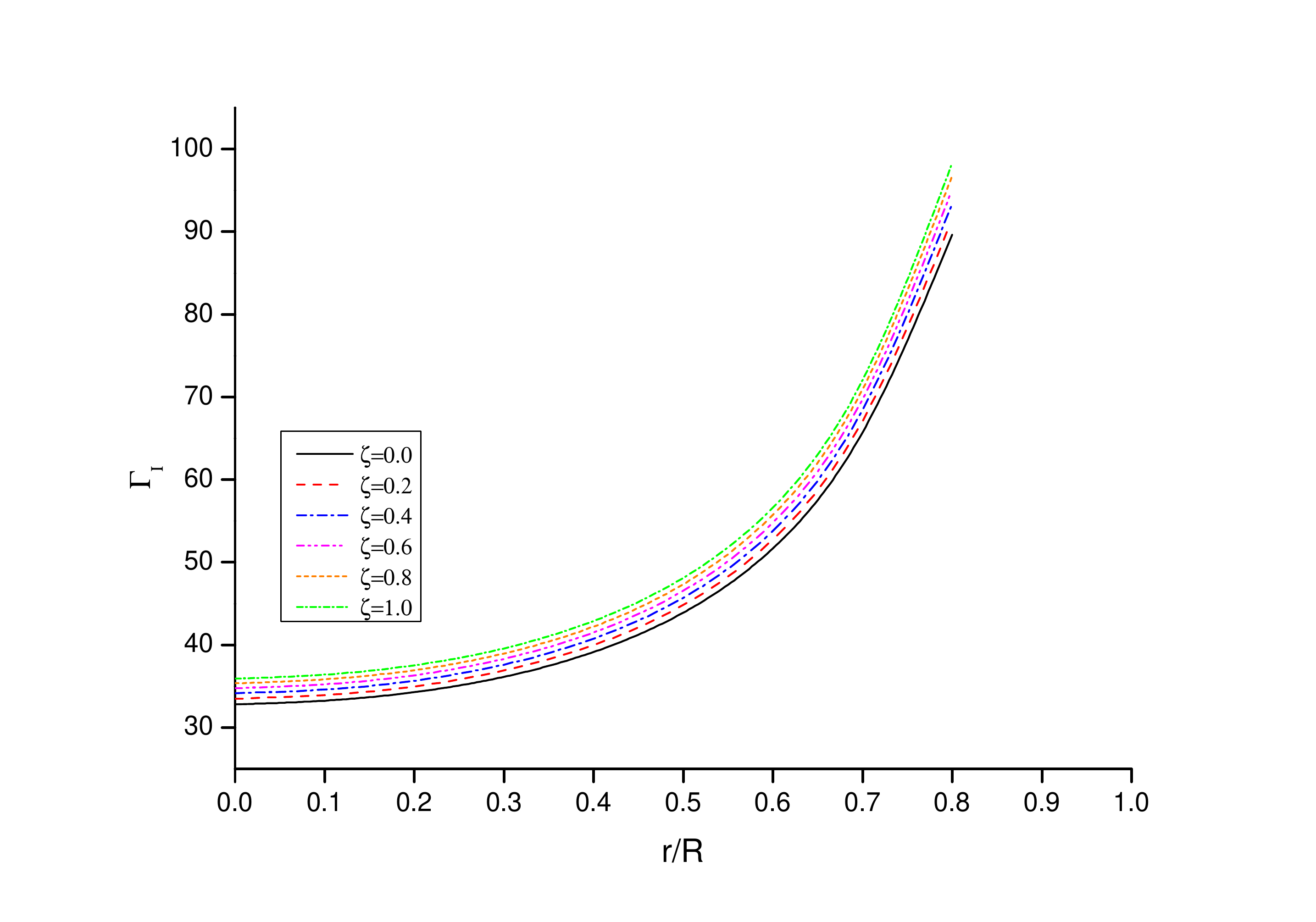}\includegraphics[width=6cm]{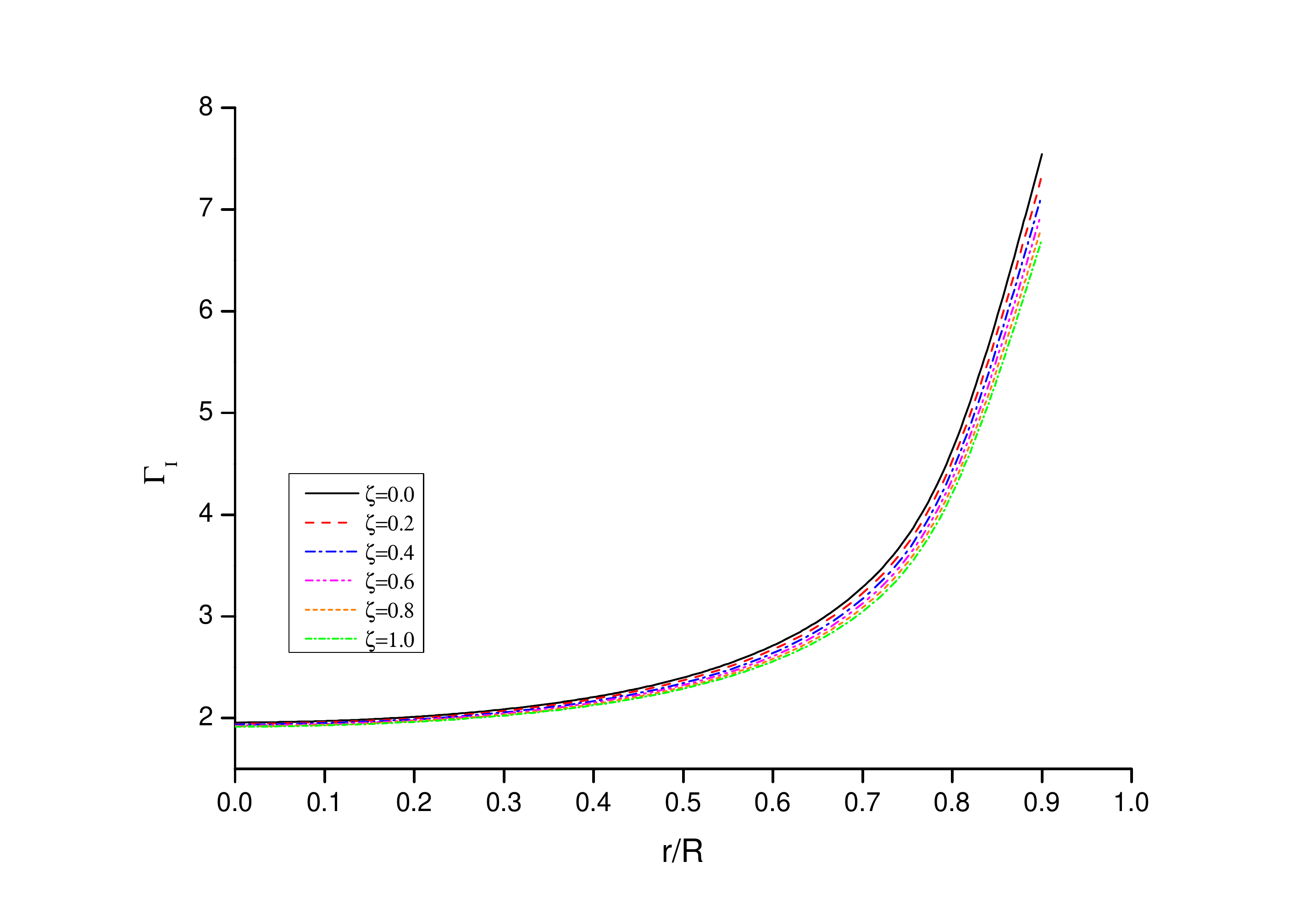}
	\caption{Behaviour of Adiabatic Index  against fractional radius $r/R$ of compact stars PSR B0943+10 and Her X-1 respectively. \label{f5}}	
\end{figure}
In {fig \ref{f5}}, we presented the Adiabatic Index $\Gamma$ of our stellar model where the value of $\Gamma >4/3$ which shows that our model has steady range with respect to the curvature term present in $f(R,T)$ gravity. But this condition may change which makes the astrophysical object more unstable due to redemptive effect of pressure.

\begin{table}[h]
	\caption{Central Pressure for different compact stars of case 1 and 2 in unit $dyne/cm^2$ \label{T1}}
	
	\begin{tabular}{cccccccc}
		\hline
		&    &  &  Case 1&  &  & &  \\
		\hline
		Compact Star&    $\varsigma=0$&  $\varsigma=0.2$&  $\varsigma=0.4$& $\varsigma=0.6$& $\varsigma=0.8$& $\varsigma=1$&  \\
		\hline
		$PSR-B0943+10$&  $1.24193 \times 10^{32}$&  $1.2039 \times 10^{32}$&  $1.16888 \times 10^{32}$& $1.13585\times 10^{32}$&  $1.10583 \times 10^{32}$&  $1.07681 \times 10^{32}$& $A_1=0$ \\
		
		CEN X-3&  $1.59908 \times 10^{34}$&  $1.61123 \times 10^{34}$&  $1.62134 \times 10^{34}$&  $1.62978 \times 10^{34}$&  $1.63656 \times 10^{34}$&  $1.6418 \times 10^{34}$& $A_1=2$  \\
		
		SMC X-4& $1.37287 \times 10^{44}$&  $1.30413 \times 10^{44}$&  $1.24138 \times 10^{44}$&  $1.18395 \times 10^{44}$&  $1.13122 \times 10^{44}$&  $1.08267 \times 10^{44}$& $A_1=-1$  \\
		
		\hline
		&  &  &  Case 2&  &  &  &   \\
		\hline
		Compact Star&    $\varsigma=0$&  $\varsigma=0.2$&  $\varsigma=0.4$& $\varsigma=0.6$& $\varsigma=0.8$& $\varsigma=1$&  \\
		\hline
		Her X-1&  $1.60709 \times 10^{44}$&  $1.60283 \times 10^{44}$&  $1.59762 \times 10^{44}$&  $1.59134 \times 10^{44}$&  $1.58443 \times 10^{44}$&  $1.57666 \times 10^{44}$&  $A_1=0$ \\
		
		$4U1538-52$&  $1.17804 \times 10^{44}$&  $1.15809 \times 10^{44}$&  $1.13919 \times 10^{44}$&  $1.12123 \times 10^{44}$&  $1.1042 \times 10^{44}$&  $1.08789 \times 10^{44}$&  $A_1=2$ \\
		\hline
	\end{tabular}
\end{table}

\begin{table}[h]
	\caption{Central Density for different compact stars of case 1 and 2 in unit $gm/cm^3$ \label{T2}}
	\begin{tabular}{cccccccc}
		\hline
		&    &  &  Case 1&  &  & &  \\
		\hline
		Compact Star&    $\varsigma=0$&  $\varsigma=0.2$&  $\varsigma=0.4$& $\varsigma=0.6$& $\varsigma=0.8$& $\varsigma=1$&  \\
		\hline
		$PSR-B0943+10$& $2.19479 \times 10^{13}$& $2.14372 \times 10^{13}$& $2.09497 \times 10^{13}$& $2.04839 \times 10^{13}$& $2.00384 \times 10^{13}$& $1.96117 \times 10^{13}$& $A_1=0$\\
		CEN X-3& $2.57064 \times 10^{14}$& $2.51209 \times 10^{14}$& $2.45623 \times 10^{14}$& $2.40287 \times 10^{14}$& $2.35185 \times 10^{14}$& $2.30299 \times 10^{14}$& $A_1=2$\\
		SMC X-4& $4.96258 \times 10^{13}$& $4.85814 \times 10^{13}$& $4.75739 \times 10^{13}$& $4.66021 \times 10^{13}$& $4.56652 \times 10^{13}$& $4.47616 \times 10^{13}$& $A_1=-1$\\
		\hline
		&  &  &  Case 2&  &  &  &   \\
		\hline
		Compact Star&    $\varsigma=0$&  $\varsigma=0.2$&  $\varsigma=0.4$& $\varsigma=0.6$& $\varsigma=0.8$& $\varsigma=1$&  \\
		\hline
		Her X-1& $7.39146 \times 10^{13}$& $7.23293\times 10^{13}$& $7.08161 \times 10^{13}$& $6.93692 \times 10^{13}$& $6.79837 \times 10^{13}$& $6.66549 \times 10^{13}$& $A_1=0$\\
		$4U1538-52$& $8.4603 \times 10^{13}$& $8.27299 \times 10^{13}$& $8.09403 \times 10^{13}$& $7.92263 \times 10^{13}$& $7.75854 \times 10^{13}$& $7.60111 \times 10^{13}$& $A_1=2$\\
		\hline
	\end{tabular}
\end{table}

\begin{table}[h]
	\centering
	\caption{Value of different parameters like $M$, $R$ and $C$ \label{T3}}	
	\begin{tabular}{cccccc}
		\hline
		&    &  Case 1&  &    \\
		\hline
		Compact Star&    $M(M_{\odot})$&  $R(Km)$&  $C(Km^{-2})$&  M/R& \\
		\hline
		$PSR-B0943+10$& 0.02& 2.6& 0.002071& 0.011415& $A_1=0$\\
		CEN X-3& 1.49& 4.178& 0.026467& 0.462008& $A_1=2$\\
		SMC X-4& 1.29& 8.831& 0.005145& 0.215479& $A_1=-1$\\
		\hline
		&  &  Case 2&  &    \\
		\hline
		Compact Star&    $M(M_{\odot})$&  $R(Km)$&  $C(Km^{-2})$&  M/R& \\
		\hline
		Her X-1& 0.85& 8.1& 0.022018& 0.154768& $A_1=0$\\
		$4U1538-52$& 0.87& 7.866& 0.024294& 0.163151& $A_1=2$\\
		\hline
	\end{tabular}
\end{table}

\clearpage
\section{Conclusion}
In this paper, we presented isotropic matter inside compact stars in the framework of modified theory of relativity i.e $f(R,T)$ gravity theory by considering Buchdahl assumptions. Extending our model to study the analytic solution for $K<0$ and $K>1$ of compact stars like PSR-B0943+10, SMC X-4, CEN X-3, Her X-1 and 4U1538-52. Futhermore, pressure inside compact stars is mostly isotropic that vanishes at the surface and density decreases with increase in radius $r$ as shown in {fig \ref{f1}}. We have shown the central density and pressure of compact stars for different parameters in {Tab \ref{T1}}. Overall here we find two solution for both the cases which are found to give equal results.  \par
\hspace{0.5cm}  At boundary $r=R$, we matched the interior with exterior Schwarzchild spacetime to find constant $\frac{c_2}{c_1}$. Applying it we find all physical quantities, we did simultaneous comparison between compact stars PSR-B0943+10 and Her X-1 at $A_1=0$ and illustrate this physical properties against fractional radius $r/R$ in Km like velocity of sound ({fig \ref{f2}}) which is less than the speed of light, energy conditions ({fig\ref{f3}}) which are non-negative, gravitational Redshift ({fig\ref{f4}}) which show satisfactory results for both stars and finally Adiabatic Index ({fig \ref{f5}}) which is greater than $4/3$. Aside this we present total mass M (normalized in solar mass $M_{\odot}$) with total radius $R$ along with different parameters like $K$ and $C$ for various compact stars in {Tab \ref{T3}} which are regular and well defined throughout the stellar structure, which shows the validity of our model. So, it can be concluded that our stellar model in reference to $f(R,T)$ Gravity Theory has admissible results that represents astrophysical objects more precisely and accurately.

\section*{Acknowledgements} 
The Authors would like to express their sincere gratitude towards Department of Mathematics, Central University of Jharkhand, Ranchi, India for the necessary support where the paper has been written and finalized.





\end{document}